\documentclass[sigconf,authorversion,nonacm]{acmart}

\AtBeginDocument{%
  }

\usepackage{xspace}
\usepackage[symbol]{footmisc}
\usepackage{amsmath}

\begin{document}

\title{Privacy Artifact ConnecTor (PACT): Embedding Enterprise Artifacts for Compliance AI Agents}

\author{Chenhao Fang*, Yanqing Peng*, Rajeev Rao*, Matt Sarmiento*, Wendy Summer*, Arya Pudota, Alex Goncalves, Jordi Mola, Hervé Robert}
\email{{chenhaofang, yanqingpeng, rrrao, mattsarmiento, wsummer, apru, alexgon, jordim, hervert}@meta.com}
\affiliation{%
  \institution{Meta}
  \city{Menlo Park}
  \state{California}
  \country{USA}
}

\renewcommand{\shortauthors}{Fang et al.}

\begin{abstract}
Enterprise environments contain a heterogeneous, rapidly growing collection of internal artifacts related to code, data, and many different tools. Critical information for assessing privacy risk and ensuring regulatory compliance is often embedded across these varied resources, each with their own arcane discovery and extraction techniques. Therefore, large-scale privacy compliance in adherence to governmental regulations requires systems to discern the interconnected nature of diverse artifacts in a common, shared universe.

We present \textbf{P}rivacy \textbf{A}rtifact \textbf{C}onnec\textbf{T}or (PACT), an embeddings-driven graph that links millions of artifacts spanning multiple artifact types generated by a variety of teams and projects.
Powered by the state-of-the-art DRAGON embedding model, PACT uses a contrastive learning objective with light fine-tuning to link artifacts via their textual components such as raw metadata, ownership specifics, and compliance context. Experimental results show that PACT's fine-tuned model improves recall@1 from 18\% to 53\%, the query match rate from 9.6\% to 69.7\% when paired with a baseline AI agent, and the hitrate@1 from 25.7\% to 44.9\% for candidate selection in a standard recommender system.

\end{abstract}

\maketitle
\footnotetext[1]{Authors contributed equally}

\section{Introduction}
\label{sec:intro}

In today's enterprise environments, AI agents are  increasingly being used to automate complex workflows~\cite{wornow2024automating}, enhance decision-making processes, and improve operational efficiency. These agents serve as assistants that can understand diverse enterprise systems, execute tasks autonomously, and provide valuable insights to human operators. To effectively perform these tasks, AI agents must navigate a vast and intricate landscape of internal artifacts, including code file repository paths, asset names, and project management tools. For example, when an AI agent is asked about a project, it must understand not only the question itself but also the context in which it is being asked, including the owner teams, code file repository paths, and tasks associated with the project. As organizations scale their operations and technological infrastructure, the demand for intelligent agents capable of managing this complex web of enterprise artifacts has grown substantially.

However, AI agents face significant challenges when attempting to navigate and operate within enterprise ecosystems. A principal challenge is how to efficiently establish connections across various types of artifacts and entities, such as linking code file repository paths to oncall teams or regulatory documents to specific products. In essence, AI agents are confronted with an intricate mapping problem whose complexity grows exponentially with the number of artifacts and entities, making it increasingly difficult for agents to establish and maintain these connections as organizations scale.

The need for a solution to this problem is especially evident in domains with strict compliance requirements, where regulations are often highly technical and precise, requiring a deep understanding of the complex web of artifacts within an organization to ensure that they are properly addressed. Conventional methods for managing artifact mappings, such as manual tracking or the use of multiple, isolated tracking systems, are no longer viable due to their limited scalability and inability to provide a unified view of artifact relationships. This limitation highlights the need for a novel approach.

In this paper, we propose using a unified embedding-based semantic graph to represent enterprise artifacts and entities and their interconnections, enabling AI agents to infer and reason about complex relationships and make informed decisions. By constructing a graph that connects various artifact types, organizations can create a powerful tool for managing enterprise artifacts and enabling AI agents to navigate complex ecosystems. We illustrate the effectiveness of this approach with a concrete use case in the domain of privacy compliance: we introduce \textbf{P}rivacy \textbf{A}rtifact \textbf{C}onnec\textbf{T}or (PACT), a privacy-focused semantic graph that connects disparate artifacts relevant to privacy compliance at Meta. This graph enables AI agents to efficiently navigate the elaborate relationships between numerous types of artifacts – including code file repository paths, teams, and projects – providing a unified view of artifact relationships for privacy management. By leveraging such a graph, organizations can better ensure compliance with regulatory requirements and improve the accuracy and timeliness of AI agent insights. PACT has been integrated into an internal AI Agent for privacy. By leveraging PACT, we achieve a more integrated and responsive approach to artifact management, ultimately supporting Meta's broader strategic objectives while maintaining compliance and operational excellence.

\section{Methodology}
In this section, we introduce the technical details of the Privacy Artifact Connector (PACT). We first describe how to build a unified semantic artifacts graph for enterprise artifacts and entities using a pre-trained embedding model in Section~\ref{sec:embedding_pretrain}. We then discuss fine-tuning the embedding model in Section~\ref{sec:embedding_ft}. The technique to build search functionality utilizing a semantic artifacts graph is discussed in Section~\ref{sec:semantic_search}. Finally, we present our method of using search functionality with an LLM-based agent in Section~\ref{sec:ai_pact}.

\subsection{Semantic Artifacts Graph Building with Pre-trained Embeddings}
\label{sec:embedding_pretrain}
Our system constructs a unified semantic graph of enterprise artifacts by first encoding each artifact into a shared embedding space. We consider a range of artifacts including but not limited to: source code file repository paths, oncall teams, products, and other internal entities that can be represented by text. For example, a code file repository can be represented by combining the file name, and an oncall team by the team name and a team charter or description. We feed these texts into the encoder of a pre-trained dense retriever -- DRAGON~\cite{lin2023traindragondiverseaugmentation} -- to obtain initial embeddings. All artifacts – irrespective of type – are projected into a single shared vector space, and their distances are negatively correlated with their semantic similarity as a result of the model's language understanding. For instance, the embedding of a code file repository and the embedding of a team will lie relatively close if they are topically related, e.g. the code file repository is owned by that team or shares nomenclature. This unified embedding space thus provides a starting point where lexical and contextual signals lead to a coarse semantic clustering of related enterprise artifacts.

\subsection{Embedding Fine-tuning}
\label{sec:embedding_ft}
We fine-tune the DRAGON embedding model on Meta internal data to sharpen and align semantic relationships. The fine-tuning process leverages known artifact links as a knowledge graph. Concretely, we construct training pairs from ground-truth relations such as code file repository path → oncall team, or team → product mappings. If artifact $A$ is linked to artifact $B^+$ in the enterprise knowledge graph (e.g. file $A$ is owned by team $B^+$), we treat $(A, B^+)$ as a positive pair. We then sample negatives by pairing $A$ with artifacts $B^-$ of the same type as $B^+$ that are not explicitly linked to $A$ (e.g. a different team for the code file repository path). The model is trained with a contrastive learning objective to distinguish true pairs from negative ones. In practice, we use an InfoNCE loss: for a given query artifact text $A$, its relevant artifacts $B+$ should have higher similarity score, while other irrelevant artifacts in the batch should be scored lower:

\[
-\log
\frac{\exp \bigl(s(A, B^{+})\bigr)}
     {\exp\bigl(s(A, B^{+})\bigr) + \sum_{j=1}^{k} \exp \bigl(s(A, B_j^{-})\bigr)}
\, .
\]
where:
\[
s(A, B) \triangleq \mathbf{e}_{A} \cdot \mathbf{e}_{B},
\]
$\mathbf{e}_{A}$ and $\mathbf{e}_{B}$ are the vectors at the last layer of embedding model.

This loss causes the encoder to adjust representations so that linked entities map close together in the vector space while unrelated ones are separated.

We also incorporate indirect relations: if $A$ → $B$ and $B$ → $C$ are known links, we consider $A$ and $C$ as semantically related. For example, a source code file repository path A can belongs to a team B which support oncall rotation C. Thus code file repository path A and oncall rotation C should be related. During training, such second-degree connections are used as additional positive pairs to propagate similarity across multi-hop chains. By iterating over these contrastive updates, the embedding model gradually encodes not just textual similarity but also the structure of the enterprise knowledge graph. After fine-tuning, artifacts connected through organizational knowledge, even artifacts that don’t share explicit keywords, end up clustered together in the embedding space. This step effectively adapts the general-purpose DRAGON model to the enterprise’s specific ontology, creating a domain-aware embedding space that integrates both content and known relationships.

\begin{figure}
    \centering
    \includegraphics[width=\linewidth]{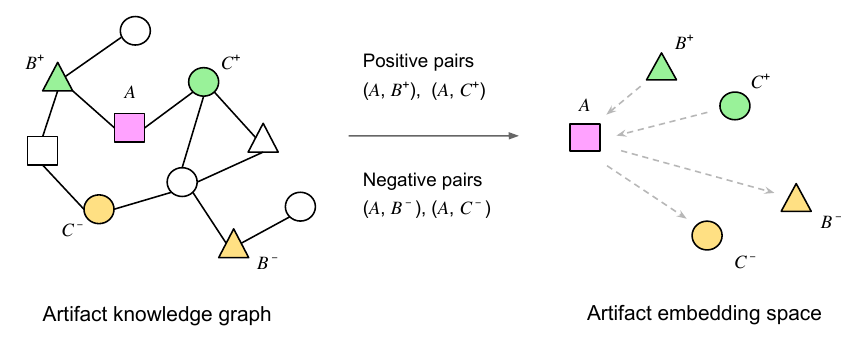}
    \caption{Embedding fine-tuning process}
    \label{fig:embed_ft}
\end{figure}

\subsection{Semantic Search System}
\label{sec:semantic_search}
Using the fine-tuned embeddings, we construct a $k$-nearest-neighbor (KNN) graph that captures the semantic relationship of all artifacts. Each artifact is a node in the graph. We add an edge between two nodes if one node belongs to the top-k nearest neighbors of the other node in the embedding space. In practice, we compute the pairwise similarities across all nodes and link each node to the $k$ most similar other nodes. The result is an undirected KNN graph $G=(V, E)$ where $V$ is the set of artifact embeddings and an edge  $(u, v)$ indicates that artifact $v$ is one of the $k$-closest neighbors of artifact $u$. This graph intrinsically connects artifacts that are semantically related, even if they are of different types. For example, a code file repository may be directly connected to the oncall team node that owns it, and also to another code file repository with similar functionality, or to a product node that the code file path is associated with. The semantic graph effectively serves as a knowledge graph over embeddings, enabling exploration of related items via graph traversal. 

To support semantic search, we utilize the unified embedding space and the KNN graph as an index. An arbitrary natural language query is handled by encoding the query string with the fine-tuned DRAGON query model to produce a query embedding vector. We then perform a nearest-neighbor lookup: the query vector is compared against the artifact embeddings to find the most similar items. Efficient approximate nearest neighbor (ANN)~\cite{arya1998optimal} search methods can be used to scale this retrieval. The top-ranked neighbors retrieved are the enterprise entities most relevant to the query’s intent. Notably, because all entity types share a single embedding space, the query is free-form: a user can enter a question or description and the system may return a mix of code file repository paths, teams, and products that are contextually appropriate. For instance, the query ``\textit{Who's the oncall for the payment service}'' might retrieve the oncall team responsible for payments, related code repositories, and documentation on payment services. The semantic graph can further enrich the search experience: once a relevant entity is found via the query, a user or downstream application can follow its graph edges to discover directly connected items, e.g., find the code file paths linked to a retrieved team. 

\begin{figure}
    \centering
    \includegraphics[width=\linewidth]{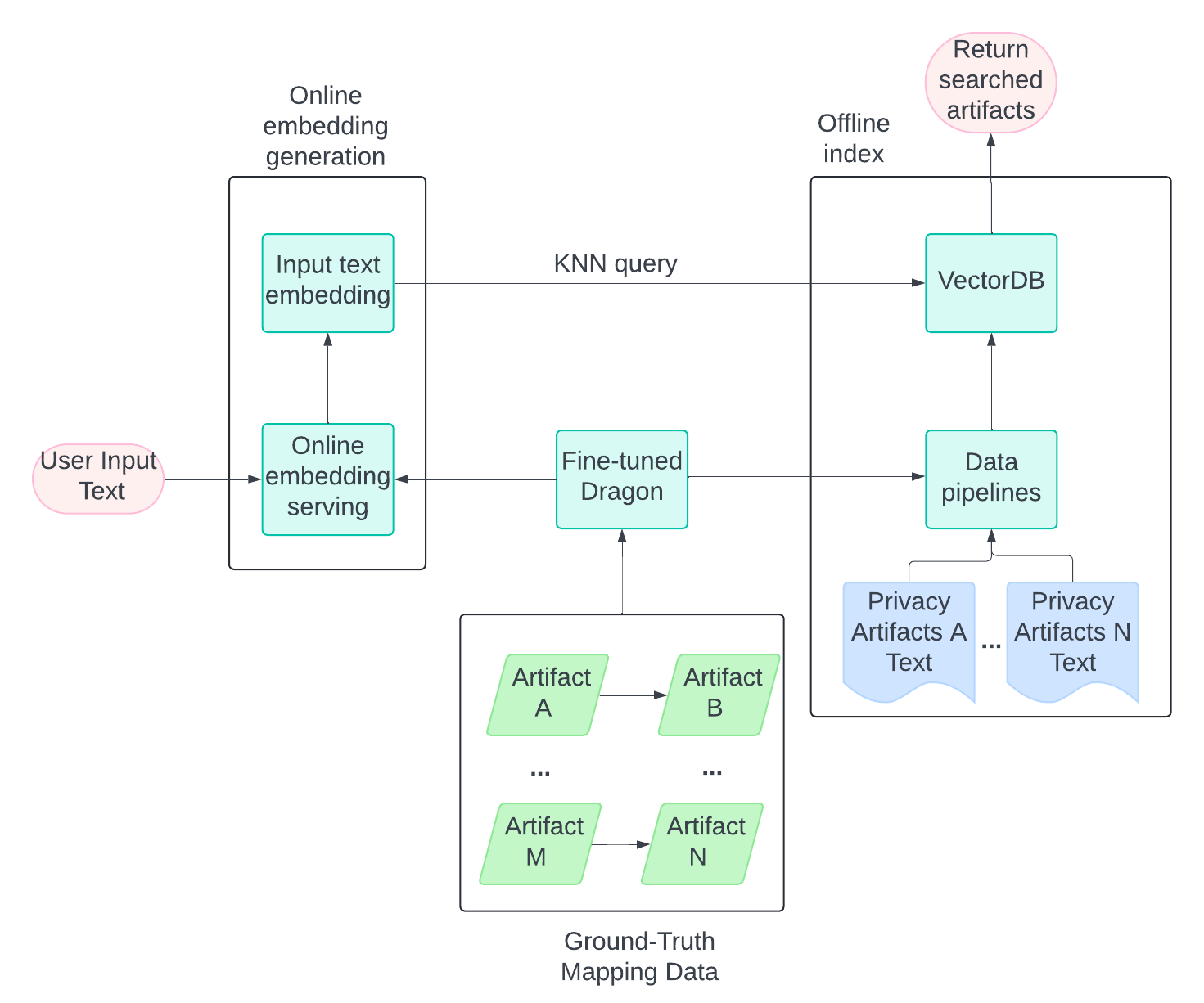}
    \caption{PACT search system}
    \label{fig:pact_architecture}
\end{figure}

Figure~\ref{fig:pact_architecture} shows the PACT search system in production environment.  
We build internal offline data pipelines to gather different textual information that best represents each artifact, and create one unified VectorDB index to store the embeddings. The VectorDB stores more than 2 million entries in total. The product quantization~\cite{jegou2010product} is also configured in the VectorDB to optimize query speed. The overall latency for the PACT system inference is fairly low, depending on the number of artifacts queried, the latency is usually from 20 ms to 200ms.

\subsection{Empowering Enterprise AI Agent with PACT}
\label{sec:ai_pact}
 We propose an LLM-based agent that is augmented with the PACT semantic text search tool outlined in Section~\ref{sec:semantic_search} to assist in reasoning over a large enterprise knowledge graph of artifacts. The agent operates in an iterative reasoning and acting loop inspired by the ReAct paradigm~\cite{yao2023react}. At each step, the agent can choose to either continue reasoning in natural language or invoke an external tool. We register the PACT-based Semantic Search tool in the agent’s action space with a defined interface: the agent can issue a search query to the tool, and in return it receives an “observation” consisting of the top-$k$ relevant artifacts. The domain-aware fine-tuned embedding model outlined in Section~\ref{sec:embedding_ft} enables it to go beyond simple keyword matching and retrieve information based on conceptual relevance, effectively giving the tool an extendable, long-term memory of enterprise knowledge. 

 \begin{figure}
    \centering
    \includegraphics[width=\linewidth]{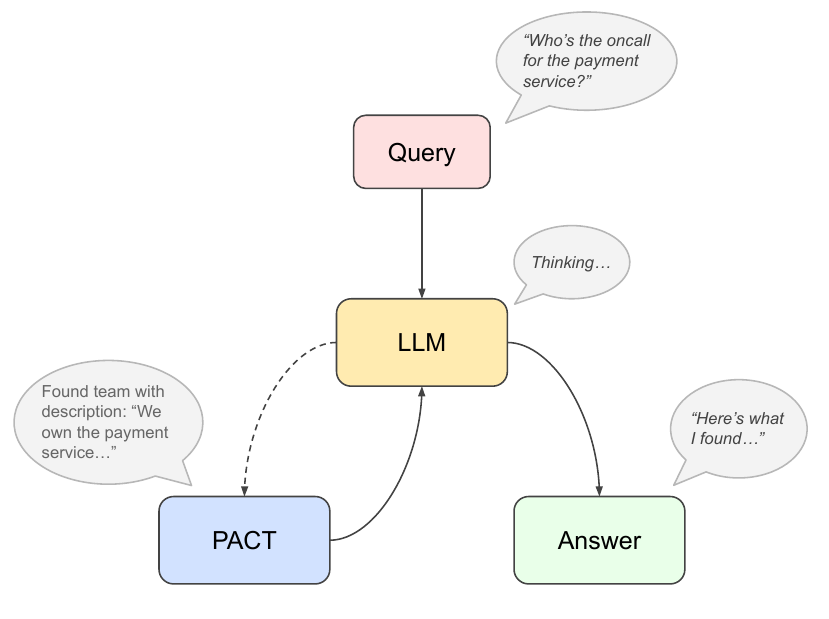}
    \caption{PACT ReAct Architecture}
    \label{fig:pact_react_architecture}
\end{figure}
 
 The agent is prompted with a description of the Semantic Search tool and guidelines for when to use it. We follow a decision-oriented prompting strategy: the system prompt includes an instruction such as \textit{``You have access to the following tools''} followed by a specification. For example, \textit{"PACT Search Tool: retrieves relevant enterprise artifacts. Use this tool to find artifacts realted to user query."} The agent’s output is expected to adhere to a format delineating thought, and observation. For example, the agent might produce: \textit{``Thought: The user’s question refers to Project Alpha. I recall seeing this name but need more details. Action: PACT Search Tool [Query: Project Alpha product].''} Upon this action, the semantic search module encodes the query and performs an approximate nearest-neighbor search in the artifact embedding index. The result is fed back to the agent as an observation,e.g., \textit{``Observation: Retrieved snippet: ‘Project Alpha is developed by Team Sky, and is a cloud analytics product...’''}. The agent then incorporates this information into its subsequent reasoning, which may lead to further tool calls or a step towards answering the query.
  
 In practice, the LLM-based agent can invoke the semantic search tool in situations below where reasoning alone is insufficient due to missing or ambiguous information. 
\subsubsection{Resolving Unknown Terms or Acronyms}
 When the task context contains an unfamiliar term, e.g. an internal project codename or an acronym, the agent will call the search tool to find definitions or relevant references. For instance, if asked \textit{``Explain how FOO module integrates with the BAR service''} and the agent does not recognize \textit{``FOO module''}, it can query the semantic index for that module’s documentation or code references to gather necessary details before proceeding with an explanation.

\subsubsection{Following Cross-References}
The enterprise artifact graph often contains implicit links – a code comment might mention a design document, or a product spec might list a team name. The agent can follow these semantic trails by issuing iterative search queries. It may start with a piece of information from one artifact and use it to find another artifact. For example, upon retrieving a design doc snippet that mentions \textit{“handled by Project Beta”}, the agent might next search for \textit{“Project Beta code repository”} or \textit{“Product Beta product page”} to understand that component, performing a multi-hop traversal. This ability to chain searches allows the agent to assemble information spread across multiple sources, which is crucial for answering holistic questions or performing root-cause analysis in complex systems.

\subsubsection{Long-Context Queries}
If a user’s query requires analyzing a large amount of information beyond the token window of the model, the agent can break the problem down and query relevant portions in stages. For example, to answer \textit{“What were the main outcomes of Project Gamma and which teams contributed to it?”}, the agent might first search for \textit{“Project Gamma product page”} to get an overview and then a follow-up search for “Project Alpha contributing teams” or search within the project report for team names. By retrieving granular pieces of the answer step-by-step, the agent effectively handles queries that exceed its immediate context length, leveraging the vector store as an extended memory.

\section{Experiments}
We conduct several experiments to validate our proposed method. The first comparison experiments is discussed in Section~\ref{sec:exp_efficacy}, in which we evaluate the effectiveness of fine-tuned embedding. In Section~\ref{sec:exp_fetcher}, we show the results of applying PACT directly as a candidate fetcher for ranking. Finally in Section~\ref{sec:exp_pact_ai_agent} We show the effectiveness of using PACT as search tool plus RAG provider for enterprise AI agent.

\subsection{Experiment 1: Efficacy of Fine-tuned Embeddings}
\label{sec:exp_efficacy}
\subsubsection{Experiments Setup}
We constructed a dataset comprising approximately 150K pairs of ground-truth data related to the privacy and compliance domain. This dataset includes information on team info, team oncalls, product hierarchies, code file repository paths at Meta. These artifacts are for internal tasks such as mapping code file paths to on-call teams, associating files with products, and clustering similar entities.
We did a $5:1$ train test split and fine-tuned an embedding model initialized from pre-trained DRAGON-RoBERTa on 8 A100 GPUs. Each batch includes a query artifact text accompanied by its positives and hard negatives pairs as required by the contrastive learning methodology. The ratio of positive and hard negative pairs was set to $1:4$ to achieve optimal information gain.

\subsubsection{Results}

\begin{figure}
    \centering
    \includegraphics[width=\linewidth]{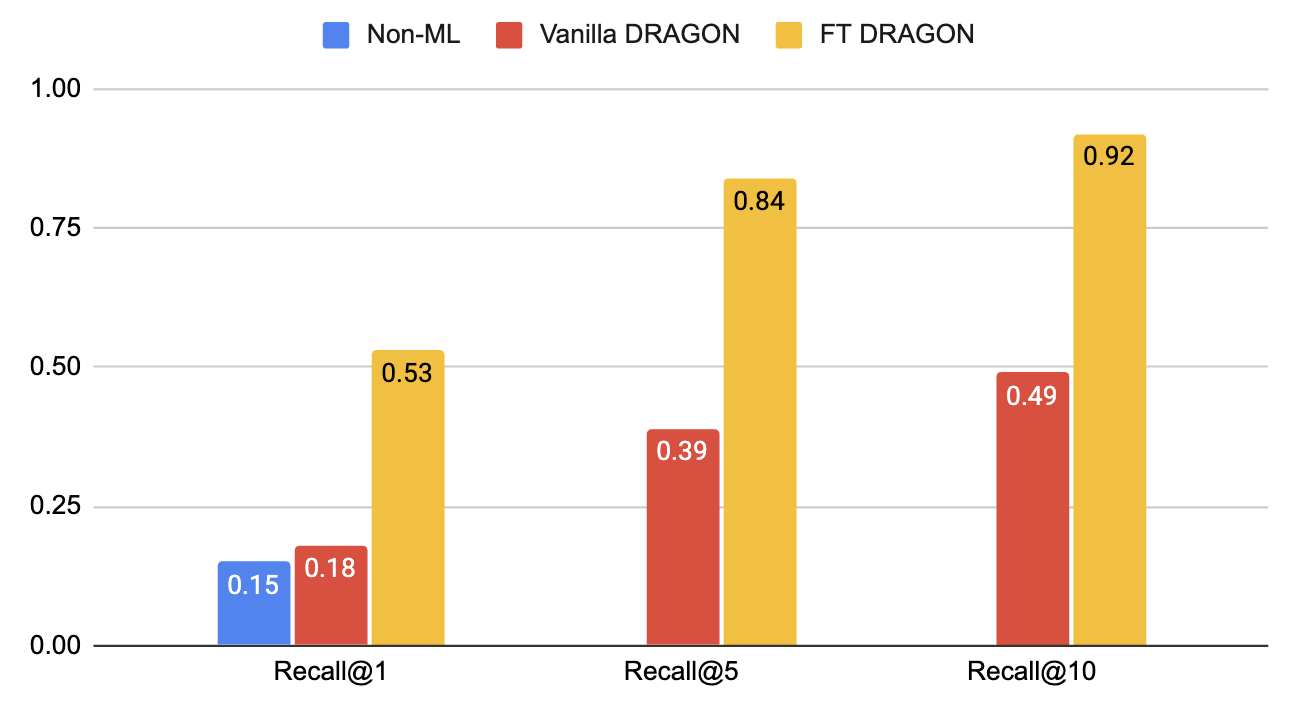}
    \caption{Comparison of recall across different models}
    \label{fig:2}
\end{figure}

We use the task of mapping code files paths to oncall teams as an example to evaluate the performance of the fine-tuned embedding model. This mapping operation is representative of many similar tasks commonplace in enterprise environments. For instance, when there is an issue that needs investigation, we want to quickly triage the problematic code to the right oncall team for quick responses. 
In the training dataset, ~50K of them correspond to the file to oncall task. The training data is carefully chosen to span most files and oncalls to maximize the information gain while also maintaining the DRAGON’s generalization ability. We use 10K production data to evaluate the performance of different methods in this task. We evaluate the recall @ 1 for human-based non-ML heuristic method, recall @ {1, 5, 10} for vanilla DRAGON-RoBERTa, and fine-tuned DRAGON-RoBERTa in the 10K test set. The results are shown in Figure~\ref{fig:2}. By injecting the knowledge and relationships between these artifacts, the quality of the model improves significantly when searching for these artifacts. 

We assess if the fine-tuning on in-domain data is causing DRAGON to overfit and lose generalization on out-of-domain data. We run the base DRAGON-RoBERTa and DRAGON fine-tuned on the standard BEIR datasets\cite{thakur2021beir}. Table~\ref{tab:1} on Normalized Discounted Cumulative Gain (NDCG) at 10 and average relevant items at top-5 show that fine-tuning doesn’t significantly degrade generalization.

\begin{table}
    \centering
    \caption{Vanilla DRAGON v.s. FT DRAGON in BEIR dataset}
    \begin{tabular}{c|c|c}
    \hline
    Embedding Models & NDCG@10 & Avg relevant items @top5\\
        \hline
      Vanilla DRAGON   & 42.0 & 65.9\\
      \hline
      DRAGON Fine-tuned   & 41.9 & 65.1\\
      \hline
    \end{tabular}
    
    \label{tab:1}
\end{table}

\subsection{Experiment 2: PACT as a Candidate Fetcher}
\label{sec:exp_fetcher}
\subsubsection{Problem Definition}
In this section, we demonstrate the use of PACT as a candidate fetcher within a standard recommender system, focusing on the challenge of accurately identifying enterprise product entities within a hierarchical tree structure. Each node in this hierarchy represents a specific product area, encompassing unique requirements and compliance considerations crucial for privacy processes. The primary task is to determine the most suitable product node for a given project based on its free-text description. This process is vital for streamlining compliance and enhancing the efficiency of internal reviews by ensuring that projects are aligned with the correct product categories and their associated compliance protocols.

\subsubsection{Methods.}

We explore three distinct methods for utilizing PACT in this task:

\textbf{Multi-Class Classification with LLM}. This straightforward approach involves presenting the project description along with all potential nodes to an LLM and prompting it to identify the correct node, similar to a multi-class classification problem. This method takes advantage of the LLM's capability to process and analyze extensive datasets.

\textbf{K-Nearest Neighbors (KNN) with PACT}. As an alternative, this approach utilizes PACT to conduct a K-Nearest Neighbors (KNN) analysis. It involves mapping the project into a feature space and identifying the closest existing nodes based on similarity metrics. By considering node proximity, this method can provide more contextually relevant suggestions, potentially enhancing accuracy compared to the simpler LLM-based classification.

\textbf{Hybrid Approach with PACT and LLM}. This method combines the strengths of both PACT and LLM. Initially, PACT serves as a candidate fetcher to narrow down the list of potential nodes, thereby reducing the complexity of the problem space. Subsequently, an LLM is used to make the final determination from the shortlisted candidates. This hybrid approach leverages PACT's precision in candidate selection and the analytical power of LLMs, resulting in improved accuracy and efficiency.

\subsubsection{Evaluation Setup}

To comprehensively evaluate our methods, we collected a subtree with approximately 350 frequently used nodes from Meta product hierarchy. Each node corresponds to a distinct product area with specific requirements and compliance considerations. We also curated a diverse dataset comprising 1,500 projects. Each project in our dataset contains metadata including the project title, detailed description, associated team information, and most importantly, the ground truth product hierarchical node manually assigned during the actual privacy review process. These ground truth labels serve as our benchmark for measuring the accuracy of each method. 

We evaluate performance using a Top-k hit rate metric (denoted as Tk), where a "hit" is recorded if the ground truth node appears within the top k recommendations. Specifically, we report T1 (exact match), T5, and T20 hit rates to assess both precision and recall capabilities. Additionally, we measure the average inference time for each method to evaluate computational efficiency.

For the Multi-Class Classification with LLM method, we employ Meta's Llama 3.3 model (70B parameter version)~\cite{grattafiori2024llama} as our LLM backbone. The model receives a prompt containing the project description and the list of product nodes, with instructions to recommend the top-k appropriate nodes. To fit the nodes within the LLM's context length, we employ a divide-and-conquer strategy. This involves recursively splitting the nodes into batches of 40 and asking the LLM to select 20 nodes from each batch in parallel. The selected nodes are then formed into a new list, and the process is repeated until only one batch remains.

For the KNN with PACT approach, we utilize embeddings generated from project descriptions using our fine-tuned text embedding model. We implement a dot product metric to identify the k-nearest nodes in the embedding space. 

For the Hybrid Approach, we first use PACT to retrieve the top-40 most similar nodes, then provide these candidates along with the project description to the LLM for final top-k recommendation. This reduces the LLM's decision space from 350 options to just 40.

\subsubsection{Results}
To evaluate the effectiveness of these methods, we conducted a series of experiments comparing their performance in identifying the correct product hierarchical nodes. The results are summarized in Table~\ref{tab:node_classification_results}.

\begin{table*}[h]
\centering
\caption{Performance and Efficiency Comparison of Product Node Classification Methods.}
\label{tab:node_classification_results}
\begin{tabular}{lcccc}
\toprule
\textbf{Method} & \textbf{T1 Hit Rate} & \textbf{T5 Hit Rate} & \textbf{T20 Hit Rate} & \textbf{Average Latency (ms)} \\
\midrule
Multi-Class Classification with LLM & 25.7\% & 45.9\% & 56.1\% & 3882 \\
KNN with PACT & 31.2\% & 47.8\% & 57.0\% & 107 \\
Hybrid Approach (PACT + LLM) & \textbf{45.0\%} & \textbf{66.8\%} & \textbf{79.8\%} & 906 \\
\bottomrule
\end{tabular}
\end{table*}

As shown in Table~\ref{tab:node_classification_results}, the Multi-Class Classification with LLM method demonstrated a baseline accuracy, providing a naive solution but often misclassifying projects in cases with subtle distinctions between nodes. We observed that the model frequently confused product nodes that had similar functionalities but different requirements.

The KNN with PACT approach showed improved accuracy over the LLM-only method, particularly in scenarios where projects had clear similarities to existing nodes. This demonstrates the effectiveness of the embedding-based semantic representation in PACT for capturing product relationships. However, it occasionally struggled with outlier projects that did not closely match any existing nodes, particularly when projects contained novel features or spanned multiple product areas.

The Hybrid Approach with PACT and LLM consistently outperformed the other two approaches, achieving the highest accuracy across all hit rate metrics. By leveraging PACT for candidate selection and LLM for final classification, this method effectively balanced precision and computational efficiency. The improvement was particularly pronounced for complex projects that required both contextual understanding (provided by the LLM) and similarity matching (provided by PACT).

We also observed that the Hybrid Approach reduced inference time by approximately 4x compared to the pure LLM approach, as the LLM needed to consider only a subset of possible nodes rather than the entire hierarchy. This efficiency gain is particularly valuable in production environments where timely classification is essential.

In conclusion, the hybrid approach of using PACT as a candidate fetcher followed by LLM classification emerges as the best method for identifying the correct product hierarchical node. 

\subsection{Experiment 3: PACT Search Tool for an Enterprise AI Agent}
\label{sec:exp_pact_ai_agent}
\subsubsection{Experiments Setup}
To evaluate the effectiveness of PACT as a search tool for an enterprise AI agent, we conducted a comprehensive assessment using a custom-crafted Artifact Understanding Benchmark dataset. This dataset consists of 54 questions that queried specific information from an enterprise knowledge graph of artifacts, with corresponding keyword ground truth answers.

The questions in the dataset were designed to test the AI agent's ability to retrieve accurate and relevant information. For instance, a sample question might ask:"\textit{Find other [artifacts of type X] related to this [artifact Y], and explain why they are related,}" with keyword ground truth answers including words like \textit{"property Z"} and other relevant keywords that highlight similar aspects.

We compared the performance of two AI agents:
\begin{itemize}
    \item Base AI agent: ReAct-style AI agent with knowledge Retrieval-Augmented Generation (RAG) capabilities, but without access to artifact tools or PACT.
    \item Base AI Agent + PACT: Base AI agent equipped with the PACT search tool described in Section~\ref{sec:ai_pact}, which enables it to retrieve and utilize specific artifacts from the enterprise knowledge graph.
\end{itemize}
By comparing the performance of these two agents on the Artifact Understanding Benchmark dataset, we aimed to assess the value added by PACT in enhancing the AI agent's ability to retrieve precise and pertinent information from the enterprise knowledge graph.
\subsubsection{Results}
In this experiment, we evaluated the performance of our models using two key metrics: 

\textbf{Average Keyword Match Rate:} Determined by calculating the match rate for each individual query and then averaging these rates.

\textbf{Global Keyword Match Rate:} Calculated by dividing the total number of keywords matched by the total number of keywords across all queries.

The results are shown in Table~\ref{tab:3}. The results of this experiment shows that the PACT tool substantially enhances the model's ability to fetch relevant information to help Agent answer artifact related questions. 

\begin{table}[h]
    \centering
    \caption{Keyword Match Rate Comparison}
    \begin{tabular}{c|c|c}
    \hline
    AI agent& Average Match Rate  & Global Match Rate\\
        \hline
      Base AI agent   & 9.6\% & 10.5\%\\
      \hline
      Agent + PACT   & 69.7\% & 60.5\%\\
      \hline
    \end{tabular}
    \label{tab:3}
\end{table}

\section{Related Work}

\subsection{Pre-trained Embedding}
Advances in language modeling have produced powerful pre-trained encoders for text, which can serve as a foundation for representing enterprise entities. BERT~\cite{devlin2019bert}, for instance, encodes textual input into a low-dimensional vector, enabling a bi-encoder retrieval setup where both queries and documents are embedded for similarity comparison. In the information retrieval community, dense retrieval models~\cite{karpukhin2020densepassageretrievalopendomain} have shown the effectiveness of task-specific fine-tuning of BERT for open-domain question answering. Building on such architectures, Lin et al. (2023) introduced DRAGON\cite{lin2023traindragondiverseaugmentation}, a BERT-based dual-encoder model trained with diverse augmentation for generalizable dense retrieval. DRAGON attains state-of-the-art retrieval accuracy on both in-domain and zero-shot evaluations without increasing model size, demonstrating that a single pre-trained embedding model can generalize well across different text domains. This model provides a strong backbone for representing varied enterprise artifacts in a unified semantic space.

\subsection{Semantic Artifact Graph}
Enterprises often maintain diverse artifacts – from source code and engineering documents to team charters and product descriptions – that can be interconnected to form a knowledge graph. Prior work has explored representing such artifacts in semantic graphs to enable unified search and analysis. For example, code graph~\cite{codegraph} have been proposed to map relationships among code entities and link them with documentation, aiding understanding of complex codebases. More generally, heterogeneous information network techniques like metapath2vec~\cite{dong2017metapath2vec} embed multi-type entities into a common vector space by leveraging meta-path guided random walks to capture multi-hop semantic relations. These approaches highlight the value of modeling connections among disparate enterprise artifacts as a graph, preserving both direct links and indirect relationships for downstream tasks such as search and recommendation.

\subsection{LLM Agent}
Large language model (LLM) agents have recently emerged as a promising direction for tackling complex tasks by extending an LLM’s capabilities with external tools. Notably, the ReAct~\cite{yao2023react} demonstrated that interleaving logical reasoning with actionable operations can significantly improve an agent’s performance on knowledge-intensive tasks. In ReAct, the model produces a chain-of-thought alongside tool calls in an interwoven manner, allowing it to gather additional information from external sources while reasoning. This approach has been shown to mitigate hallucinations and error propagation by grounding the reasoning process in retrieved facts. Subsequent works expanded on the idea of equipping LLMs with tool-use abilities. For example, Toolformer \cite{schick2023toolformerlanguagemodelsteach} introduced a method to fine-tune LLMs to decide when and how to invoke external APIs such as search engines or calculators during generation. By training on self-supervised traces of API calls, Toolformer imbues a language model with the autonomy to call tools for factual lookup or computations, yielding improved zero-shot performance on tasks requiring external knowledge. 

\section{Conclusion}

We present \textbf{P}rivacy \textbf{A}rtifact \textbf{C}onnec\textbf{T}or (PACT), an embeddings-driven graph that connects disparate artifacts in an enterprise environment. PACT's primary idea combines a contrastive learning objective with light fine-tuning to co-locate multiple artifacts with similar textual information in common neighborhoods in a multi-dimensional embedding space. Generating these linkages allows us to discover the interconnected nature of a diverse entities and serves as a foundational element for large-scale Privacy compliance. We describe a real-world application where PACT is used as the preeminent tool for an LLM agent to serve as the provider of Search results as well as RAG information to enhance downstream prompts. Further, we illustrate an enterprise search solution where the combination of PACT as a candidate fetcher and an LLM as a ranker produces significant improvement on both efficiency and effectiveness over generic search alternatives.

\begin{acks}
We sincerely thank Xilun Chen, Reese Wynn, Jan-Willem Maessen, Hengli Liang and Vaibhav Shrivastava for their help in building the system, and Mark Harman for his
valuable feedback and thoughtful discussions throughout this work.
\end{acks}

\bibliography{acmart}

\end{document}